\documentclass[aps,pre,twocolumn,amsmath,amssymb,superscriptaddress,floatfix,showpacs]{revtex4}
\usepackage{graphicx}
\usepackage{bm}
\bibstyle{apsrev.bib}

\newcommand{\be}{\begin{equation}}
\newcommand{\ee}{\end{equation}}
\newcommand{\bea}{\begin{eqnarray}}
\newcommand{\eea}{\end{eqnarray}}

\begin{document}

\title{Aperiodic Ising model on the Bethe lattice: Exact results}

\author{Ferenc Igl\'oi}%
 \email{igloi@szfki.hu}
 \affiliation{Research Institute for Solid State Physics and Optics,
H-1525 Budapest, P.O.Box 49, Hungary}
 \affiliation{Institute of Theoretical Physics,
Szeged University, H-6720 Szeged, Hungary}
 \affiliation{Laboratoire de Physique des Mat\'eriaux, Nancy-Universit\'e, CNRS, BP~239,\\
F-54506 Vand\oe uvre l\`es Nancy Cedex, France}
\author{Lo\"{\i}c Turban}
 \email{turban@lpm.uhp-nancy.fr}
 \affiliation{Laboratoire de Physique des Mat\'eriaux, Nancy-Universit\'e, CNRS, BP~239,\\
F-54506 Vand\oe uvre l\`es Nancy Cedex, France}

\date{\today}

\begin{abstract}
We consider the Ising model on the Bethe lattice with aperiodic modulation of the
couplings, which has been studied numerically in Phys. Rev. E {\bf 77}, 041113 (2008).
Here we present a relevance-irrelevance criterion and solve the critical behavior
exactly for marginal aperiodic sequences. We present analytical formulas for the
continuously varying critical exponents and discuss a relationship with the (surface)
critical behavior of the aperiodic quantum Ising chain.
\end{abstract}


\maketitle

\section{Introduction}
\label{sec:1}
Disorder and different types of inhomogeneities are inevitable features of
real materials. Their presence may modify the physical properties of a system and their
effect can be particularly strong close to singularities, such as at phase transition points \cite{maccoy73,cardy96}.
In some cases the perturbation can change the universality class of a second-order phase
transition. In this respect relevance or irrelevance of an inhomogeneous perturbation
can be analyzed in terms of linear stability at the pure system's fixed point as first 
performed by Harris \cite{harris74} for uncorrelated bond disorder. The classification
of the critical behaviors of disordered systems with a random fixed point is a challenging and
theoretically very difficult task.

Another type of inhomogeneities is introduced by quasiperiodic or, more generally, aperiodic
modulations of the couplings. Since the discovery of quasicrystals \cite{shechtman84} and due to the progress 
in molecular beam epitaxy, allowing for the preparation of good quality
multilayers with a prescribed aperiodic structure \cite{majkrzak91}, there has been an increased interest
to study theoretically the phase transitions in such nonperiodic systems \cite{luck94}. These 
systems can be considered as somehow intermediate between pure and random ones and are
expected to display a rich variety of critical behaviors. Indeed a generalization of
the Harris criterion predicts that, depending on the strength of the fluctuations of the aperiodic sequence and the value of the correlation length critical exponent $\nu$ of the pure system, an aperiodic
perturbation may be irrelevant, marginal, or relevant \cite{luck93,igloi93}. A series of  
works on the critical behavior in different aperiodic systems  confirms 
the validity of the generalized relevance-irrelevance criterion \cite{godreche86,okabe88,okabe90,sakamoto89,henley87,turban94a,turban94b,berche95,berche96,igloi97a,hermisson97,hermisson98,vieira05}.

Interestingly in the presence of aperiodicity one can observe truly marginal behavior. Then
the critical exponents are nonuniversal and their value varies continuously with the amplitude of the aperiodic perturbation. Such a behavior has been obtained exactly for the aperiodic Ising quantum chains using a renormalization group transformation or a finite-size scaling analysis \cite{turban94a,turban94b,berche96,igloi97a,hermisson97,hermisson98}, as well as for the
interface delocalization transition in the Penrose quasiperiodic lattice \cite{henley87}. Nonuniversal critical behavior is expected to occur in real higher-dimensional systems, too, for example, in a three-dimensional tricritical system where $\nu=1/2$. In this case, however, no exact results are available yet. Numerical studies of the related mean-field model
with Fibonacci modulation of the couplings show nonuniversal critical behavior \cite{igloi97b,berche97}.

More recently the Bethe-lattice Ising model, which also belongs to the mean-field universality class \cite{baxter}, has been studied numerically for two types of perturbations \cite{faria08}. With a Fibonacci modulation of the couplings, the classical mean-field exponents are recovered  whereas, for a period-doubling (PD) modulation, the magnetic exponents are nonuniversal. The difference in
the relevance of the Fibonacci modulation for the two mean-field models is due to the
different ways in which the mean-field behavior is realized, but this question has not been studied so far.

In this paper we continue the study of the aperiodic Bethe-lattice Ising model. Our motivations are twofold: First, we are interested in
the formulation of a relevance-irrelevance criterion adapted to this system in order to explain
the conflicting results of previous numerical works. Second, and our more important motivation, we can provide an {\it exact solution} of the problem and in this way we obtain analytical
formulas for the continuously varying critical exponents, among others for the PD
sequence studied before numerically.

The structure of the paper is the following. The model, the aperiodic sequences and the
corresponding relevance-irrelevance criterion are presented in Sec. \ref{sec:2}. The critical
behavior of marginal aperiodic sequences is studied analytically in Sec. \ref{sec:3}, in which
we point out a close relationship with the (surface) critical properties of aperiodic
quantum Ising chains. Our results are discussed in Sec. \ref{sec:4} and details about
the calculation of sums of aperiodic variables are given in the Appendix.

\section{Aperiodic perturbation and its relevance}
\label{sec:2}
\subsection{Hamiltonian}
%%%%%%%%%%% FIG 1  %%%%%%%%%%%%%%%%%%%%%%%%%%%%%%%

\begin{figure}
\begin{center}
\includegraphics[width=\columnwidth]{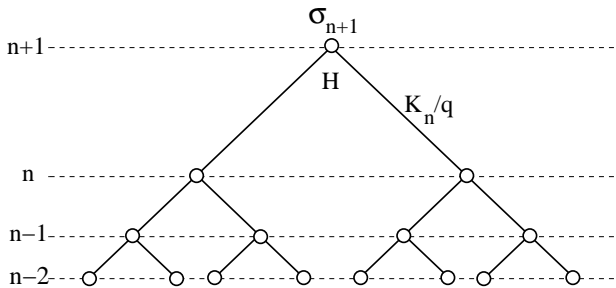}
\end{center}
\caption{Aperiodic Bethe lattice with coordination number $z=q+1=3$. The couplings $K_n/q$ are aperiodically modulated.}
\label{fig1}
\end{figure}

%%%%%%%%%%%%%%%%%%%%%%%%%%%%%%%%%%%%%%%%%%%%%%

We consider the spin 1/2 Ising model on a Bethe lattice with coordination number $z=q+1$. The Hamiltonian reads
\be
-\beta{\cal H}=\sum_n\sum_{(i,j)\in n} \frac{K_n}{q}\sigma_i\sigma_j+H\sum_i\sigma_i\,,
\label{e2-1}
\ee
where $\beta=1/k_{\mathrm B}T$ is the inverse temperature. The first sum runs over the successive layers of the lattice indexed by $n$ (see Fig. \ref{fig1}) and the second over the bonds between the sites in layers $n+1$ and $n$. The couplings $K_n/q$ are aperiodically modulated and properly normalized in order to allow us to take the mean-field limit $q\to\infty$. They are parametrized as
\be
K_n=Kr^{f_n}\,,\qquad f_n=0,1\,,
\label{e2-2}
\ee
where $r$ is the ratio of perturbed to unperturbed couplings and the binary variables $f_n$ follow some aperiodic sequence.

Let ${\cal O}_n=\langle{\cal O}_i\rangle$ be the thermal average of some local operator ${\cal O}_i$ in layer $n$. This average is fluctuating from layer to layer due to the aperiodic modulation of the couplings. Following Ref. \cite{faria08}, a mean value $\overline{{\cal O}}$ is defined by giving the same weight to the different layers,
\be
\overline{{\cal O}}=\frac{1}{N}\sum_{n=1}^N {\cal O}_n\,.
\label{e2-3}
\ee 
With this choice, the usual mean-field critical behavior is obtained for the unperturbed system.
When each layer is weighted according to its number of spins, the surface spins dominate the critical behavior which is then quite unusual \cite{mullerhartmann74}: There is  no long-range order at $T>0$ and the free energy displays a power-law singularity in $H$ with a critical exponent increasing smoothly from 1 to infinity as $T$ goes from 0 to the Bethe-Peierls temperature. 

The Bethe lattice may be embedded in a $d=\infty$ hypercubic lattice by placing each link into a different lattice direction. Then the geometrical distance $L$, measured in lattice parameter units, between two spins located $N$ layers apart grows as \cite{saw}
\be
L^2=N
\label{e2-4}
\ee
since the $L$ steps are mutually orthogonal.

\subsection{Aperiodic sequences}
As in Ref. \cite{faria08} we consider aperiodic sequences generated via substitutions on the binary digits 0 and 1. For the PD sequence \cite{collet80} we have ${\cal S}(0)=0\ 1$ and ${\cal S}(1)=0\ 0$ which, starting on 0, give successively
\bea
&&0\,,\nonumber\\
&&0\ 1\,,\nonumber\\
&&0\ 1\ 0\ 0\,,\nonumber\\
&&0\ 1\ 0\ 0\ 0\ 1\ 0\ 1\,.
\label{e2-5}
\eea
The properties of the sequence can be deduced from its substitution matrix \cite{queffelec87,dumont90}
with entries $M_{ij}$ giving the numbers $n_i^{{\cal S}(j)}$ of digits of type $i$ in ${\cal S}(j)$. In the case of the PD sequence one obtains
\be
M=\left(\begin{array}{cc}
n_0^{{\cal S}(0)}&n_0^{{\cal S}(1)}\\
n_1^{{\cal S}(0)}&n_1^{{\cal S}(1)}
\end{array}
\right)
=\left(\begin{array}{cc}
1&2\\
1&0
\end{array}
\right)\,.
\label{e2-6}
\ee
The entries in $M^p$ give the numbers of digits of each type in the sequence after $p$ iterations. The length $N$ of the sequence obtained after $p$ iterations (which is also the number of layers on the Bethe lattice) is related to the leading eigenvalue $\Lambda_1$ of the substitution matrix through $N=\Lambda_1^p$. Let 
\be
n_N=\sum_{n=1}^N f_n\,,\qquad \rho_\infty=\lim_{N\to\infty}\frac{n_N}{N}\,,
\label{e2-7}
\ee
be the number of 1 in a sequence with $N$ digits and the corresponding asymptotic density, respectively. 
On the Bethe lattice, according to Eq. (\ref{e2-3}), the mean value of the coupling is given by
\bea
\overline{K}&=&\lim_{N\to\infty}\frac{1}{N}\sum_{n=1}^NK_n=K+\lim_{N\to\infty}\frac{n_N}{N}K(r-1)\nonumber\\
&=&K+\rho_\infty \Delta\,,
\label{e2-8}
\eea
where $\Delta=K(r-1)$ is the amplitude of the aperiodic modulation of the couplings.
The mean deviation from $\overline{K}$ on a system with $N\sim\Lambda_1^p$ layers takes the form
\bea
\overline{\delta K}(N)&=&\frac{1}{N}\sum_{n=1}^N(K_n-\overline{K})
=\Delta\left(\frac{n_N}{N}-\rho_\infty\right)\nonumber\\
&\sim&\frac{\Delta}{N}\Lambda_2^p\sim\Delta N^{\omega-1}\,,
\label{e2-9}
\eea
where $\omega$ is the wandering exponent of the sequence given by
\be
\omega=\frac{\ln |\Lambda_2|}{\ln \Lambda_1}\,,
\label{e2-10}
\ee
in terms of the second leading eigenvalue $\Lambda_2$ of the substitution matrix.   
For the PD sequence, according to Eq. (\ref{e2-6}), we have $\Lambda_1=2$, $\Lambda_2=-1$, so that $\omega=0$.

\subsection{Relevance-irrelevance criterion}
The Harris argument \cite{harris74}, showing that thermal randomness is a relevant perturbation only when the specific heat exponent $\alpha$ of the pure system is positive, has been generalized to the case of aperiodic perturbations in Refs. \cite{luck93,igloi93}. 

The argument can be adapted to our problem as follows: Near the critical point of the pure system, the relevant length is the correlation length $\xi$ diverging as $t^{-\nu}$ where $t\sim|K-K_c|$ measures the deviation from the critical temperature. To the length $\xi$ is associated the number of layers $N=\xi^2$ according to Eq. (\ref{e2-4}) and the aperiodic perturbation induces a shift in the critical temperature 
$\overline{\delta t}\sim\overline{\delta K}(N)$ which, according to Eq. (\ref{e2-9}), takes the form 
$\overline{\delta t}\sim \xi^{2(\omega-1)}\sim t^{-2\nu(\omega-1)}$. The ratio 
\be
\frac{\overline{\delta t}}{t}\sim t^{-\phi}\,,\qquad \phi=1+2\nu(\omega-1)
\label{e2-11}
\ee
gives the relative strength of the aperiodic perturbation. It diverges, and thus the perturbation is relevant, when the crossover exponent $\phi>0$. It is irrelevant when $\phi<0$ and marginal when $\phi=0$. In this latter case the aperiodicity may lead to a nonuniversal behavior with some exponents varying continuously with the amplitude of the perturbation.

The same result can be obtain by studying the scaling behavior of the perturbation amplitude $\Delta$ in Eq. (\ref{e2-9}). Under a change of the length scale by a factor $b=L/L'$, Eq. (\ref{e2-4}) leads to $N'=N/b^2$ 
and $\overline{\delta K}(N)$, with scaling dimension $y_t=1/\nu$, and transforms as
\be
(\overline{\delta K})'\sim \Delta' {N'}^{\omega-1}=b^y_t\Delta N^{\omega-1}
\label{e2-12}
\ee
so that
\be
\Delta'=b^{y_t+2(\omega-1)}\Delta\,.
\label{e2-13}
\ee
Thus the scaling dimension of $\Delta$ is $\phi/\nu$ and the perturbation grows under rescaling (is relevant) when $\phi$ is positive. 

A continuous  variation of the magnetic exponents was observed in Ref. \cite{faria08} for the PD sequence with $\omega=0$. This marginal behavior is expected since $y_t=1/\nu=2$ for the mean-field Ising model.
On the contrary, the Fibonacci sequence, with $\omega=-1$ \cite{turban94b}, leads to an irrelevant perturbation. It does not change the critical behavior which remains classical.

\section{Critical behavior}
\label{sec:3}

\subsection{Finite-size behavior of the magnetization}
We consider an $n+1$-generation branch defined as an initial site with spin $\sigma_{n+1}$ connected to $q=z-1$ $n$-generation branches as shown in Fig. \ref{fig1}; a one-generation branch is a single site. Let $Z_n^\pm$ be the sum of the contributions to the partition function of an $n$-generation branch with initial spin either up ($+$) or down ($-$). It satisfies the recursion relation
\be 
Z_{n+1}^\pm=e^{\pm H}\left(e^{\pm K_n/q} Z_n^+ +e^{\mp K_n/q} Z_n^-\right)^q\,.
\label{e3-1}
\ee
In the mean-field limit $q\to\infty$, the $n$th layer magnetization $m_n=\langle\sigma_n\rangle$ may be written as
\be 
m_n=\frac{Z_n^+ -Z_n^-}{Z_n^+ +Z_n^-}
\label{e3-2}
\ee
since the contribution of the single branch going forward can be neglected compared to the contributions of the $q$ branches going backward. Expanding the exponentials in Eq. (\ref{e3-1}) one obtains
\be
Z_{n+1}^\pm\!\!=\!e^{\pm H}\!\!\left[Z_n^+ \!+\!Z_n^- \!\pm\!\frac{K_n}{q} (Z_n^+\! -\!Z_n^-)\!+\!O\left(\!\frac{K_n^2}{q^2}\!\right)\right]^q
\label{e3-3}
\ee
so that
\bea
\lim_{q\to\infty}\frac{Z_{n+1}^\pm}{(Z_n^+ +Z_n^-)^q}
&=&e^{\pm H}\lim_{q\to\infty}\left[1\pm\frac{K_n}{q}m_n
+O\left(\frac{K_n^2}{q^2}\right)\right]^q\nonumber\\
&=&\exp[\pm(H+K_nm_n)]\,,
\label{e3-4}
\eea
and the layer magnetization satisfies the recursion relation
\be
m_{n+1}=\tanh(H+K_nm_n)\,.
\label{e3-5}
\ee
Expanding to the first order in the external field $H$ and to the third order in the first layer magnetization $m_1$, one has
\bea
m_{n+1}&=&H\prod_{i=1}^nK_i\sum_{k=1}^n\prod_{j=1}^kK_j^{-1}+m_1\prod_{i=1}^nK_i\nonumber\\ 
&&\ \ \ -\frac{m_1^3}{3}\prod_{i=1}^nK_i\sum_{k=1}^n\prod_{j=1}^kK_j^2+\cdots\,.
\label{e3-6}
\eea
According to Eq. (\ref{e2-3}) the mean value of the magnetization on a system with $n$ layers is given by
\be
\overline{m}=\frac{1}{N}\sum_{n=1}^Nm_n=a_0\,H+a_1\,m_1-\frac{a_3}{3}\,m_1^3+\cdots
\label{e3-7}
\ee
with
\bea
a_0&=&\frac{1}{N}\sum_{n=1}^{N-1}\prod_{i=1}^nK_i\sum_{k=1}^n\prod_{j=1}^kK_j^{-1}\,,\nonumber\\
a_1&=&\frac{1}{N}\left(1+\sum_{n=1}^{N-1}\prod_{i=1}^{n}K_i\right)\,,\nonumber\\
a_3&=&\frac{1}{N}\sum_{n=1}^{N-1}\prod_{i=1}^nK_i\sum_{k=1}^n\prod_{j=1}^kK_j^2\,.
\label{e3-8}
\eea
The critical point of the system is obtained by analyzing the asymptotic behavior of $a_1$.
It is divergent (goes to zero), if $\left(\prod_{j=1}^N K_j\right)^{1/N}$ is greater (smaller) than 1.
Consequently the critical point is given by the condition
\be
\lim_{N \to \infty} \frac{1}{N} \sum_{j=1}^{N} \ln K_j = \overline{\ln K}=0\;,
\ee
leading to 
\be
K_c=r^{-\rho_\infty}\,.
\label{e3-9}
\ee.

\subsection{Relation with the one-dimensional Ising model in a transverse field}
Let us consider the inhomogeneous quantum Ising chain with Hamiltonian
\be
H=-\sum_l J_l s_l^zs_{l+1}^z-h\sum_ls_l^x\,,
\ee
where $s_l^x$ and $s_l^z$ are the components of a Pauli spin operator associated with site $l$,  $J_l$ is the first-neighbor exchange interaction, and $h$ is the transverse field. On a chain with size $L$ and the end spin fixed, the surface magnetization $m_s$ satisfies the relation \cite{peschel84}
\be
m_s^{-2}=1+\sum_{l=1}^{L}\prod_{i=1}^{l}\lambda_i^{-2}\,,
\ee
where $\lambda_i=J_i/h$. This is just the form of the sum giving $Na_1$ in Eq. (\ref{e3-8}). At the critical coupling, generally given by $\overline{\ln \lambda}=0$ \cite{pfeuty79},  $m_s^{-2}$ scales like $L^{2x_s}$ where $x_s=\beta_s/\nu$ is the scaling dimension of the surface magnetization with $x_s=1/2$ for the unperturbed quantum Ising chain.

This exponent has been determined analytically in the case of  marginal aperiodic modulations of the couplings \cite{turban94a,turban94b,berche96,igloi97a}. Since $N$ is replaced by $L$,  $2\nu=1$ is replaced by $\nu_{\rm Ising}=1$ in the expression (\ref{e2-11}) of the crossover exponent $\phi$. Thus marginal behavior is obtained for the same value $\omega=0$ of the wandering exponent. 

With the parametrization $\lambda_i=\lambda r^{f_i}$, one obtains 
\be
x_s(r)=\frac{\ln(r^{1/3}+r^{-1/3})}{2\ln 2}
\label{e3-10}
\ee
for the PD sequence. Changing $r$ into $r^{-1}$ does not change $x_s$ for this sequence, but this is not generally true. 

The leading behaviors of the sums appearing in Eq. (\ref{e3-8}) are calculated in the Appendix.  One may notice that since $K_i$ corresponds to $\lambda_i^{-2}$, $r$ in Eq. (\ref{e3-10}) has to be replaced by $r^{-1/2}$ in the scaling exponent of $a_1$. At the critical point, the different coefficients  scale with $N$ as follows
\bea
a_0&\!\sim\!&N^{x_0}\,,\quad x_0=2x_s(r^{1/2})+2x_s(r^{-1/2})-1\,,\nonumber\\
a_1&\!\sim\!&N^{x_1}\,,\quad x_1=2x_s(r^{-1/2})-1\,,\nonumber\\
a_3&\!\sim\!&N^{x_3}\,,\quad x_3=2x_s(r^{-1})+2x_s(r^{-1/2})-1\,.
\label{e3-11}
\eea

\subsection{Finite-size scaling and critical exponents}
%%%%%%%%%%% FIG 2  %%%%%%%%%%%%%%%%%%%%%%%%%%%%%%%

\begin{figure}
\begin{center}
\includegraphics[width=.9\columnwidth]{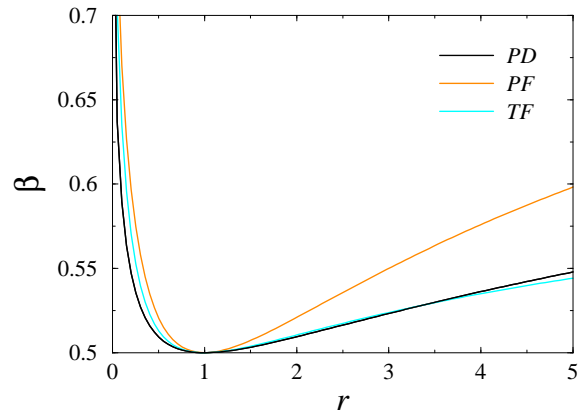}
\end{center}
\vglue -10mm
\caption{(Color online) Variation of the critical exponent $\beta$ of the spontaneous magnetization with the coupling ratio $r$ for the PD, paper-folding (PF), and three-folding (TF) sequences. The exponent is minimum and takes its mean-field value $\beta=1/2$ for the unperturbed system at $r=1$.}
\label{fig2}
\end{figure}

%%%%%%%%%%%%%%%%%%%%%%%%%%%%%%%%%%%%%%%%%%%%%%
%%%%%%%%%%% FIG 3  %%%%%%%%%%%%%%%%%%%%%%%%%%%%%%%

\begin{figure}
\begin{center}
\includegraphics[width=.9\columnwidth]{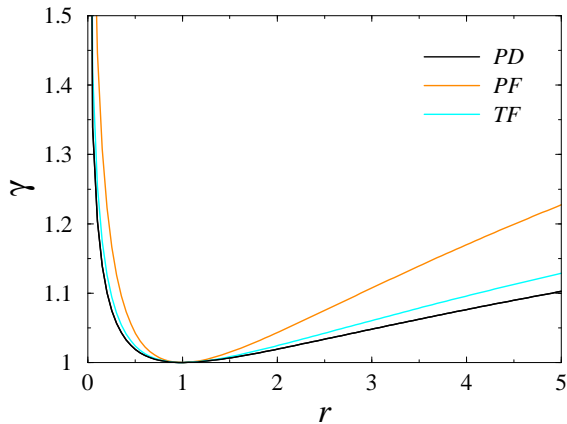}
\end{center}
\vglue -10mm
\caption{(Color online) Variation of the critical exponent $\gamma$ of the susceptibility with the coupling ratio $r$ for the PD, PF, and TF sequences. The exponent is minimum and takes its mean-field value $\gamma=1$ for the unperturbed system at $r=1$.}
\label{fig3}
\end{figure}

%%%%%%%%%%%%%%%%%%%%%%%%%%%%%%%%%%%%%%%%%%%%%%

In a finite-size system with $N$ layers the free energy density $\overline{g}$ is a function of the deviation from the critical temperature  $t$ and the external field $H$, of the system size $N$ and also of the magnetization $m_1$ of the first layer. Under a change of the length scale by a factor $b$, these variables transforms as
\be
t'=b^{y_t}t\,,\  H'=b^{y_H}H\,,\  N'=\frac{N}{b^2}\,,\  m_1'=b^{2\tau}m_1\,.
\label{e3-12}
\ee
For a truly marginal system, the thermal dimension $y_t=1/\nu$ has to keep its unperturbed value $y_t=2$. 

The critical exponents $\beta$, governing the temperature dependence of the spontaneous magnetization, and $\gamma$, governing the temperature dependence of the susceptibility, have the following expressions
\be
\beta=\frac{d_c-y_H}{y_t}\,,\quad \gamma=\frac{2y_H-d_c}{y_t}\,,\quad y_t=2\,,
\label{e3-18}
\ee 
where $d_c$ is the upper critical dimension of the problem. The dimensions $y_H$, $\tau$, and $d_c$ remain to be determined for the aperiodic system.
 
The free energy density transforms as
\be
\overline{g}(t,H,N,m_1)=b^{-d_c}\overline{g}\left(b^{y_t}t,b^{y_H}H,\frac{N}{b^2},b^{2\tau}m_1\right)\,.
\label{e3-13}
\ee
The mean value of the magnetization $\overline{m}=\partial\overline{g}/\partial H$ reads
\be
\overline{m}(t,H,N,m_1)=b^{-d_c+y_H}\overline{m}\left(b^{y_t}t,b^{y_H}H,\frac{N}{b^2},b^{2\tau}m_1\right)\,.
\label{e3-14}
\ee
It can be expanded in powers of $m_1$ as
\bea
\overline{m}(t,H,N,m_1)&=&\sum_{k\geq0}b^{y_H-d_c+2k\tau}\nonumber\\
&&\times\overline{m}^{(k)}\left(b^{y_t}t,b^{y_H}H,\frac{N}{b^2}\right)m_1^k\,.
\label{e3-15}
\eea
At the critical temperature, the leading dependence on a small external field H comes from the term of order zero in the expansion (\ref{e3-15}). With $t=0$ and $b^2=N$ we have
\bea
\overline{m}(0,H,N,m_1)&=&N^{(y_H-d_c)/2}\varphi_H^{(0)}(N^{y_H/2}H)+O(m_1)\nonumber\\
&\sim&N^{(2y_H-d_c)/2}H \,.
\label{e3-16}
\eea
When $H=0$, $\overline{m}$ is odd in $m_1$ so that, with $b^2=N$, one obtains
\bea
\overline{m}(t,0,N,m_1)\!\!&=&\!\!N^{\tau+(y_H-d_c)/2}\varphi_t^{(1)}(Nt)m_1\nonumber\\
&&\!\!\!\!\!\!\!\!+N^{3\tau+(y_H-d_c)/2}\varphi_t^{(3)}(Nt)m_1^3+\cdots\,.
\label{e3-17}
\eea
%%%%%%%%%%% TABLE 1 %%%%%%%%%%%%%%%%%%%%%%%%%%%%%%

\begin{table}
\caption{Critical behavior of the Bethe lattice with an aperiodic modulation following the PD sequence: comparison of the exact values of the critical exponents $\beta$, $\gamma$ in Eq. (\protect\ref{e3-19}) and $\delta$ deduced from the Widom scaling law to the numerical values obtained in Ref. \protect\cite{faria08}.\label{tab1}}
\vskip 0.2truecm
\begin{ruledtabular}
\begin{tabular}{llll}
&$r=1$&$r=2$&$r=7$ 
\\
\hline
$\beta$ (exact)&1/2&0.5094795 &0.5675859  \\
$\beta$ (numerical)&&0.5093(4)&0.5664(5)\\
$\gamma$ (exact)&1&1.0192114 &1.1491583  \\
$\gamma$ (numerical)&&1.0197(2)&1.1499(4)\\
$\delta$ (exact)&3&3.0004955 &3.0246421  \\
$\delta$ (numerical)&&3.0006(2)&3.0266(9)\\
\end{tabular}
\end{ruledtabular}
\end{table}

%%%%%%%%%%%%%%%%%%%%%%%%%%%%%%%%%%%%%%%%%%%%%%
Comparing Eqs. (\ref{e3-16}) and (\ref{e3-17}) to Eqs. (\ref{e3-7}) and (\ref{e3-11}) one can deduce the values of the critical exponents
\bea
\gamma&=&x_0=2x_s(r^{1/2})+2x_s(r^{-1/2})-1\,,\nonumber\\
\beta&=&\frac{x_3-3x_1}{2}=x_s(r^{-1})-2x_s(r^{-1/2})+1\,,\nonumber\\
\tau&=&\frac{x_3-x_1}{2}=x_s(r^{-1})\,,\nonumber\\
d_c&=&2+4x_s(r^{-1})+4x_s(r^{1/2})-4x_s(r^{-1/2})\,.
\label{e3-19}
\eea 
For the unperturbed system ($r=1$, $x_s=1/2$) the mean-field Ising values, $\gamma=1$,  $\beta=1/2$, $d_c=4$, are recovered.
The variations of $\beta$ and $\gamma$ with $r$ for the PD sequence  are shown in Figs. \ref{fig2} and \ref{fig3}. Similar results for the paper-folding (PF) and three-folding (TF) sequences are also shown \cite{seq}. In these cases  the values \cite{berche96,igloi97a} 
\be
x_s(r)=\frac{\ln(1+r^{-1})}{2\ln 2}\ {\rm (PF)}\,,\quad x_s(r)=\frac{\ln(2+r)}{2\ln 3}\ {\rm (TF)}\,.
\label{e3-20}
\ee
have been used in Eq. (\ref{e3-19}).

For the PD sequence, the values of $\beta$, $\gamma$, and $\delta$ (deduced from the Widom scaling law $\delta=1+\gamma/\beta$) given in Table \ref{tab1} for $r=2$ and $r=7$ compare  well with the numerical values obtained in Ref. \cite{faria08} (notice that $r$ in the present work corresponds to $r-1$ in \cite{faria08}).

\section{Discussion}
\label{sec:4}

In this paper we have studied the critical behavior of the Ising model on a Bethe lattice
with aperiodic modulation of the couplings. Our first result is the
relevance-irrelevance criterion of Eq. (\ref{e2-11}) which is adapted to the Bethe-lattice problem. As for the aperiodic quantum Ising chain or previous mean-field models, it has a form typical of a one-dimensional aperiodicity. The difference lies in the fact that the length of the sequence does not scale here like a physical length $L$ but like the number of layers $N$ on the Bethe lattice, which itself scales like $L^2$. As a consequence, the correlation length exponent $\nu$ is replaced by $2\nu$. Thus for the Bethe-lattice problem with $\nu=1/2$ the aperiodicity is marginal when the wandering exponent $\omega=0$ as for the Ising quantum chain with $\nu_{\rm Ising}=1$. For the same reason, the aperiodicity is irrelevant on the Bethe lattice for the Fibonacci sequence with $\omega=-1$ \cite{faria08} whereas it is marginal for the one-dimensional mean-field models \cite{igloi97b,berche97}.

We have solved the critical properties of the aperiodic Bethe-lattice Ising model exactly and
we have observed further similarities with the quantum Ising chain. For marginal aperiodic sequences, such as the PD sequence, the critical exponents are nonuniversal in both cases  and
the Bethe-lattice exponents can be expressed in terms of the surface magnetization exponent 
of the quantum Ising chain, taken at different values of the aperiodic coupling ratio $r$.
Since for the quantum Ising chain there is a vast literature about exact solutions for different marginal sequences, from these we can immediately translate the corresponding analytical results for the Bethe lattice.

We have also noticed that for the Bethe lattice, in order to satisfy the scaling relations, a
varying upper critical dimension $d_c(r)$ has to be introduced. As a matter of fact,
this result follows from an analysis of the Ginzburg criterion \cite{ginzburg60}, too. Something similar occurs  for the Ising quantum chain, in which the dynamical exponent  $z$ 
was found to be  $r$ dependent \cite{berche95,berche96}. As a consequence, here also the
effective dimension of the system  $d=1+z$  varies continuously with $r$.

Our investigation could be extended into several directions. For relevant aperiodic sequences,
such as the Rudin-Shapiro sequence, first-order transition is expected in one range of the
ratio, $r<1$, whereas in the other range, $r>1$, the magnetization should exhibit an essential
singularity at the critical point \cite{igloi94}, instead of a power law observed for marginal perturbations.
This type of essential singular behavior is probably the rule for random interactions, too.
\begin{acknowledgments}
  This work has been
  supported by the Hungarian National Research Fund under Grants No. OTKA
TO48721, No. K62588, No. K75324, and No. MO45596. 
 The Laboratoire de Physique des Mat\'eriaux is Unit\'e
  Mixte de Recherche CNRS No. 7556.
\end{acknowledgments}

\appendix*

\section{Calculation of the sums through renormalization}
Let us consider the  sum
\be
S_N(K,r)=1+\sum_{p=1}^N\prod_{i=1}^pKr^{f_i}=\sum_{p=0}^NK^pr^{n_p}\,,\  n_0=0\,,
\label{a-1}
\ee 
such that $a_1$ in Eq. (\ref{e3-8}) is given by $S_{N-1}(K,r)/N$.
For the PD sequence the following relations are satisfied (see Ref. \cite{turban94a} where 0 and 1 are exchanged)
\be
f_{2k}=1-f_k\,,\  f_{2k+1}=0\,,\  n_{2k}=n_{2k+1}=k-n_k\,.
\label{a-2}
\ee
Splitting the sum into even and odd parts and ignoring minor end corrections, one obtains
\bea
S_N(K,r)&=&\sum_{k=0}^{N/2}K^{2k}r^{n_{2k}}+\sum_{k=0}^{N/2}K^{2k+1}r^{n_{2k+1}}\nonumber\\
&=&\sum_{k=0}^{N/2}(K^2r)^kr^{-n_{k}}+K\sum_{k=0}^{N/2}(K^2r)^kr^{-n_{k}}\nonumber\\
&=&(1+K)S_{N/2}(K^2r,r^{-1})\,.
\label{a-3}
\eea
A second iteration leads to
\be
S_N(K,r)=(1+K)(1+K^2r)S_{N/4}(K^4r,r)
\label{a-4}
\ee
which is a renormalization transformation leaving $r$ invariant, dividing $N$ by 4 and changing 
$K$ into $K'=K^4r$. In the infinite system this transformation has a nontrivial fixed point at
\be
K^*=K_c=r^{-1/3}
\label{a-5}
\ee
which is the critical coupling of the problem, in agreement with the general result of Eq. (\ref{e3-9}). 
At the critical point, Eq. (A4) gives the finite-size
scaling relation
\bea
S_N(K_c,r)&=&(1+r^{-1/3})(1+r^{1/3})S_{N/4}(K_c,r)\nonumber\\
&=&(r^{1/6}+r^{-1/6})S_{N/4}(K_c,r)\,.
\label{a-6}
\eea
Injecting the power law $S_N(K_c,r)\simeq AN^{\omega(r)}$ into Eq. (A6), one finally obtains
\be
\omega(r)=\frac{2\ln(r^{1/6}+r^{-1/6})}{2\ln 2}=2x_s(r^{-1/2})
\label{a-7}
\ee
in agreement with the value of $x_1$ given in Eq. (\ref{e3-11}).

The values of $Na_0$ and $Na_3$ in Eq. (\ref{e3-8}) are given by the sum
\be
T_N(K,r;m)=\sum_{p=1}^NK^pr^{n_p}\sum_{l=1}^pK^{ml}r^{mn_l}
\label{a-8}
\ee
with $m=-1$ for $a_0$ and $m=2$ for $a_3$. The leading contribution to this sum 
for large $N$ values is given by
\bea
T_N(K,r;m)&\simeq& \sum_{p=0}^NK^pr^{n_p}S_p(K^m,r^m)\nonumber\\
&\simeq& A\sum_{p=0}^NK^pr^{n_p}p^{\omega(r^m)}\,.
\label{a-9}
\eea
Proceeding as before for $S_N$, the sum can be split into even and odd parts and after two iterations one obtains \cite{remark}
\be
T_N(K_c,r;m)\simeq 4^{\omega(r^m)}(r^{1/6}+r^{-1/6})T_{N/4}(K_c,r;m)\,.
\label{a-10}
\ee
It follows that, at the critical point, this sum scales with $N$ as
\bea
T_N(K_c,r;m)&\sim& N^{\omega(r;m)}\,,\quad \omega(r;m)=\omega(r)+\omega(r^m)\nonumber\\
\omega(r;m)&=&2x_s(r^{-1/2})+2x_s(r^{-m/2})\,.
\label{a-11}
\eea
With the appropriate values of $m$ one recovers the exponents in Eq. (\ref{e3-11}).


\begin{thebibliography}{99}
\bibitem{saw} Once the Bethe lattice has been embedded in the hypercubic lattice, the path between two spins belonging to different layers is a $N$-step self-avoiding walk and the end-to-end distance $L$ is given by Eq. (\protect\ref{e2-4}).
\bibitem{seq} The PF sequence results from the two-digit substitutions ${\cal S}(00)=10\ 00$ ${\cal S}(01)=10\ 01$ ${\cal S}(10)=11\ 00$ ${\cal S}(11)=11\ 01$. In this case the substitution matrix is constructed by considering the substitutions on the four letters $A=00$, $B=01$, $C=10$, and $D=11$. 
The TF sequence is more simply obtained via the substitutions ${\cal S}(0)=0\ 1\ 0$ and ${\cal S}(1)=0\ 1\ 1$.
\bibitem{remark} In this calculation the factor $(2k+1)^\omega$ is replaced by $(2k)^\omega$, which does not change the leading contribution to the sum.
\end{thebibliography}
\end{document}